\documentclass[a4paper,10pt,twoside]{cpc-hepnp}
\usepackage{multicol}
\usepackage{graphicx}
\usepackage{booktabs}
\usepackage{amssymb,bm,mathrsfs,bbm,amscd}
\usepackage[tbtags]{amsmath}
\usepackage{lastpage}
\usepackage{CJK}
\usepackage{graphicx}

\usepackage[scientific-notation=true]{siunitx}
\usepackage{amsmath}
\usepackage{amsfonts}
\usepackage{amssymb}
\usepackage{multirow}
\usepackage{tikz}
\usetikzlibrary{shapes.geometric, arrows}

\usepackage[utf8]{inputenc}
\usepackage[colorlinks=true, linkcolor=red, citecolor=blue]{hyperref}

\def\({\left(}
\def\){\right)}
\def\[{\left[}
\def\]{\right]}

\def\be{\begin{eqnarray}}
\def\ee{\end{eqnarray}}
\usepackage[capitalise]{cleveref}
\crefname{figure}{Fig.}{Figs.}
\Crefname{figure}{Fig.}{Figs.}
\def\hubble{~\mathrm{km~s^{-1}~Mpc^{-1}}}

\begin{document}

\fancyhead[c]{\small Chinese Physics C~~~Vol. **, No. ** (**)
**} \fancyfoot[C]{\small **-\thepage}

\footnotetext[0]{Received \today}

\title{The prospects of gravitational waves on constraining the anisotropy of the Universe\thanks{Z.C. and Z.-C.Z. are supported by grants from NSFC ( 11675182 and 11690022 ).  H.-N.L. are supported by grants from NSFC ( 11603005 and 11775038 ).}}

\author{%
      Zhi-Chao Zhao$^{1,2;1)}$\email{zhaozc@ihep.ac.cn}%
\quad Hai-Nan Lin$^{3;2)}$\email{linhn@cqu.edu.cn}%
\quad Zhe Chang$^{1,2;3)}$\email{changz@ihep.ac.cn}
}
\maketitle

\address{%
$^1$ Institute of High Energy Physics, Chinese Academy of Sciences, Beijing 100049, China\\
$^2$ School of Physical Sciences, University of Chinese Academy of Sciences, Beijing 100049, China\\
$^3$ Department of Physics, Chongqing University, Chongqing 401331, China\\
}

\begin{abstract}
The observation of GW150914 indicated a new independent measurement of the luminosity distance of a gravitational wave event.  In this paper, we constrain the anisotropy of the Universe by using gravitational wave events. We simulate hundreds of events of binary neutron star merging that may be observed by Einstein Telescope. Full simulation for producing process of gravitational wave data is employed. We find that 200 of binary neutron star merging in redshift $(0,1)$ observed by Einstein Telescope may constrain the anisotropy with an accuracy comparable to the result from Union2.1 supernovae. This result shows that gravitational waves can be a powerful tool in investigating the cosmological anisotropy.
\end{abstract}

\begin{keyword}
gravitational wave, anisotropic space-time, gravitational wave detector
\end{keyword}

\begin{pacs}
98.70.Vc, 98.80.Cq, 04.30.-w
\end{pacs}

\footnotetext[0]{\hspace*{-3mm}\raisebox{0.3ex}{$\scriptstyle\copyright$}2013
Chinese Physical Society and the Institute of High Energy Physics
of the Chinese Academy of Sciences and the Institute
of Modern Physics of the Chinese Academy of Sciences and IOP Publishing Ltd}%

\begin{multicols}{2}

\section{Introduction}
\label{sec:intro} 
\noindent
The cosmological principle implies that the Universe is homogeneous and isotropic on large scales. This assumption has been verified by many observations, such as the statistical properties of galaxies distribution~\cite{TrujilloGomez:2010yh} and of the cosmic microwave background radiation from the Wilkinson Microwave Anisotropy Probe ({WMAP})~\cite{Larson:2010gs,Bennett:2012zja} and Planck satellites~\cite{Ade:2013zuv,Ade:2015xua,Aghanim:2018eyx}. Based on this principle, $\Lambda$ Cold Dark Matter model ({$\Lambda$CDM}) has been proposed and it is in good agreement with many experiments. However, some other observations indicate possible confliction with the cosmological principle, such as the anisotropy of the cosmic microwave background radiation spectrum~\cite{Ade:2013zuv, Ade:2015xua}, inconsistency of the fine structure constants in the observations of the north and south celestial spheres~\cite{Murphy:2003hw,Webb:2010hc,King:2012id,Pinho:2016mkm}, and the anisotropy of distance redshift relationship given by type-Ia supernovae~\cite{Antoniou:2010gw,Mariano:2012wx,Kalus:2012zu}. These observations strongly suggest that there may exist tiny anisotropy in the Universe. If the Universe is really anisotropic, there should be a new physics beyond the standard cosmological model.

It is a natural way to study cosmological anisotropy by the relationship of luminosity distance and redshift. 
The recently observed gravitational waves (GW)~\cite{Abbott:2016blz,TheLIGOScientific:2017qsa} provide us a new way of observing luminosity distances independent of cosmological models.
The GW is self-calibrating. The luminosity distance of the GW source can be obtained just from the GW signal. If the GW signal has an electromagnetic counterpart, the redshift of the GW source can be measured by the electromagnetic counterpart. 

GW170817, a breakthrough in the history of GW observation, is a signal of a binary neutron star (BNS) merging. Its electromagnetic counterparts showed the source located at NGC4993 ($z\sim0.01$). The luminosity distance is given by the GW signal~\cite{TheLIGOScientific:2017qsa}. aLIGO\&Virgo collaborations used these data to constrain Hubble Constant and gave the result of $H_0 = 70.0^{+ 12.0}_{- 8.0}\hubble$~\cite{Abbott:2017xzu}. It implies that the GW is a very powerful tool in constraining cosmology parameters. 
In this paper, we study the possibility to constrain the anisotropy of space-time by using the GW.

We notice that compared with the detectors expected to be built later, aLIGO\&Virgo is not a powerful one~\cite{Moore:2014lga}. It can only detect the BNS merging event of about $z<0.05$~\cite{TheLIGOScientific:2014jea,Camp:2015saa}. Thus we consider the third generation detector of GW, Einstein Telescope (ET)~\cite{Punturo:2010zz}. ET is a GW detector still under conceiving, which will have three arms at an angle of 60 degrees to each other, providing the ability to detect signals at $z \sim 1$ for BNS merging events, as a possible design~\cite{amaro2009einstein}. The main purpose of this paper is to investigate the possibility of constraining the anisotropy of the Universe by using ET.

It should be noted that Fisher Matrix has been used to constrain cosmological parameters~\cite{2011PhRvD..83b3005Z,Cai:2017aea,Lin:2018azu,Chang:2019xcb}. There is no doubt that Fisher Matrix is a powerful tool in reaching parameters' uncertainty of a model with future observations. It can give the lower limit of the uncertainty of the model parameters. 
However the lower limit of the uncertainty given by Fisher Matrix can only be achieved when all system and instrumental errors are considered and best data processing is used. In the actual GW detection, data processing is very complicated. One can not guarantee to get the lower limit of the uncertainty given by Fisher Matrix.
In this paper, we use a new way to predict the uncertainty. We get the GW waveform by simulating the BNS merging events, and use the GW waveform and the noise of the GW detectors to obtain simulated signals. Markov-Chain Monte-Carlo ({MCMC}) is employed to infer the cosmological parameters from these simulated signals. This procedure completely simulates the real GW signal processing. 
We will use the results of Planck 2015 as fiducial parameters for the standard cosmology model~\cite{Ade:2015xua}. Bilby~\cite{Ashton:2018jfp} is employed as the signal injection and parameter inference tool.

The rest of this paper is organized as follows. In Section \ref{sec:method}, we simulate GW signals and make parameter inferences to obtain luminosity distance uncertainties. In Section \ref{sec:constraint}, We constrain the anisotropy of the Universe by making use of the simulated GW. Section \ref{sec:conclusion} is devoted to concluding remarks.

\section{Cosmological models and $\sigma_{d_L} \sim d_L$ curve}
\label{sec:method}
\noindent

Gravitational waves are fluctuations of space-time metric. 
Standard Siren is a GW event that both GW and corresponding electromagnetic signal are observed. The observed GW170817 is such a BNS merging event~\cite{TheLIGOScientific:2017qsa}. 

The GW generated by the BNS merging event at inspiral stage can be roughly described by the post-Newton (PN) approximation. Here we use the TaylorF2 model, which is also currently used by aLIGO\&Virgo~\cite{Buonanno:2009zt}. The TaylorF2 model is a purely analytic PN model. It includes point-particle and aligned-spin terms to 3.5PN order as well as leading order (5PN) and next-to-leading-order (6PN) tidal effects~\cite{Boyle:2009dg}. There are several parameters in this model that affect the waveform of GW propagating to earth: the masses, spins, tidal deformability parameters of the two stars, the luminosity distance of the source, the inclination angle between the two stars angular momentum and our sight, the sky position of the source, the time that GW propagates to geocentric, and the initial phase and the polarization angle.

For the sake of convenience, here we take values for the parameters: 

\begin{itemize}
	
	\item[-] The mass of the two neutron stars is $1.4M_{\odot}$. $1.4M_{\odot}$ is a typical neutron star mass.
	
	\item[-] We assume the neutron stars to be spinless. Because we usually consider that stars in a binary system are old neutron stars. If the period of spin is much larger than the period of revolution, the spin effect can be ignored. Works suggest that radio observable pulsars in BNS have a distribution of spin periods that extends down to about 15 ms~\cite{Tichy:2012rp,Oslowski:2009zr,Kiel:2008xw}.
	
	\item[-] We assume that the angular momentum of the BNS is in line with our line of sight. This is because the electromagnetic radiation generated in the event of the binary neutron star merger is along the direction of the orbital momentum. If the direction is inconsistent with our line of sight, we will unable to observe the electromagnetic signal~\cite{Nakar:2007yr}.
	
	\item[-]  We assume that of the GW events are homogeneously distributed on the celestial sphere.
	
	\item[-] The luminosity distance is described by the fiducial $\Lambda$CDM.
	
	\item[-] Assuming that the phase angle is homogeneously distributed in (0, 2$\pi$).
	
	\item[-] The polarization angles are homogeneously distributed in (0,$\pi$).
	
	\item[-] Assuming a homogeneous distribution of arrival times.
	
	\item[-] The dimensionless tidal deformability parameter of the two stars is dependent on the equation of state of neutron star. However, this parameter has little effect on our result. Referring to Figure 5 of Ref.~\cite{TheLIGOScientific:2017qsa}, we take 425 for both neutron stars.
	
\end{itemize}


We add the GW signal to the detector noise and then employ MCMC to infer the parameters. In parameter inference, we fix the masses, spins, angular momentum direction, sky position of the source (can be accurately observed with electromagnetic signals) and tidal deformability parameters. Only the remaining four parameters need to be inferred, i.e. $\(d_L, \phi, \varphi, t_0\)$, luminosity distance, polarization angle, phase and arrival time.
We use Bilby~\cite{Ashton:2018jfp} for simulation and parameter inference, use dynesty\footnote{\url{https://github.com/joshspeagle/dynesty}} as the MCMC sampler and set npoints = 5000, dlogz = 0.01.

Ref.~\cite{amaro2009einstein} showed that the horizon for BNS merging event of ET is expected to reach redshifts of $z\sim1$. We choose 10 points in $z=\(0,1\)$ with step size $\Delta z=0.02$. For each $z$, we simulate 160 BNS merging events.
In each simulation, we use the standard cosmological model to calculate the luminosity distance, homogeneously take the direction ($l,b$), phase, polarization angle, arrival time according to the method described in the previous section. Then we employ MCMC to infer the luminosity distance, polarization angle, phase and arrival time. In parameters inferencing, we set the prior of $d_L$ distributing uniformly in comoving volume from $0$ to $5d_{L_\text{inject}}$, where $d_{L_\text{inject}}$ is the injected luminosity distance. In the simulation process, only ``GW detectable" BNS events are considered. We deem BNS events “GW detectable” when the network of ET has signal-to-noise ratio (SNR)$>8$.

For a certain redshift $z$, we only use the inferred uncertainty of $d_L$. There is a center value and two uncertainties ${d_L}^{- \sigma^-_{d_L}}_{+ \sigma^+_{d_L}}~$ in the result.
For convenience, we combine them to $$\sigma_{d_L}' = \sqrt{\frac{(\sigma_{d_L}^+)^2 + (\sigma_{d_L}^-)^2}{2}}~.$$ Thus we get 160(For high redshfit, due to some of their SNR $< 8$, the number will be less than 160) $\sigma_{d_L}'$, the average of the 160 $\sigma_{d_L}'$ are
\begin{equation}\label{sigmadl}
\sigma_{d_L}^2 = \frac{1}{\frac{1}{160} \sum_{i=1}^{160}\left(\frac{1}{\left(\sigma^{'}_{{d_L}^i}\right)^2} \right) }~.
\end{equation}

%

We do the same procedure for all redshifts $z$, and a $\sigma_{d_L}/d_L \sim d_L$ curve is got. 
We use second-order polynomials to fit the curve and obtain
\begin{align}\label{ET_fitted}
\frac{\sigma_{d_L} }{ d_L} &= -1.15 \times 10^{- 9} d_L^2 + 2.53 \times 10^{- 5} d_L~.
\end{align}
The $\sigma_{d_L}/d_L \sim d_L$ curve is ploted in Fig.~\ref{etuncertainty}

\begin{center}
	\includegraphics[width=\columnwidth]{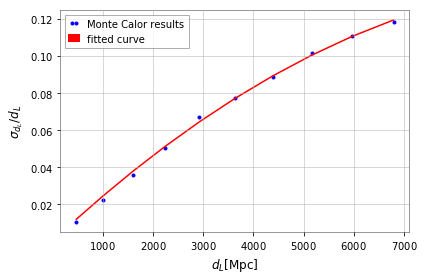}
	\figcaption{\label{etuncertainty}   The fitting formula of $\sigma_{d_L}/{d_L} $ as a function of redshift $d_L$. The blue dots denote the results based on Monte Carlo sampling, while the red solid curve denotes a fitting formula in Eq.~(\ref{ET_fitted}).}
\end{center}

\section{Constraining on the Anisotropy}
\label{sec:constraint}

In this section, firstly we introduce the anisotropic model of the cosmology.
The anisotropic model possesses more parameters than $\Lambda$CDM, one for anisotropic amplitude and two for anisotropic direction.
Then we use the $\sigma_{d_L}/d_L \sim d_L$ curve obtained in the previous section to prospect the uncertainties of the anisotropy parameters in the future.

The anisotropic model of cosmology can be described as a simple dipole model of the distance modulus. It has been discussed by Ref.~\cite{Lin:2018azu,Lin:2016jqp}. The distance modulus of this model is of the form
\be\label{modelA}
\mu = \mu_{\Lambda \text{CDM}} (1 - d \cos \theta)~,
\ee
where
\be\label{distancemodulus}
\mu = 5 \log \frac { d _ { L } } { \mathrm { Mpc } } + 25~,
\ee
and $\theta$ is the  angle between the direction of space-time anisotropy and the direction of the event. The direction of space-time anisotropy can be parameterized as ($l,b$) in the
galactic coordinate system.  $d$ is the anisotropic amplitude. $\mu_{\Lambda \text{CDM}}$ is the distance modulus in fiducial $\Lambda$CDM model.
The fiducial parameters in this model is given by the result of Union 2.1~\cite{Lin:2016jqp}, which are $d = 0.0012, l=310.6^\circ, b=-13^\circ$.


Next, we use the curve of Eq. (\ref{ET_fitted}) to simulate BNS merging events to be observed by ET and infer the relevant parameters ($d, l, b$) in the anisotropic model of cosmology. We do our simulation as follows. For example, for the case of 100 observed BNS merging events:

\begin{enumerate}
	\item The redshift of each event is referred according to the event rate ($z\in\(0,1\)$)\cite{2011PhRvD..83b3005Z,Lin:2018azu}. After taking the time evolution of the burst rate into account, we can write the event rate in this form,
	\be\label{eventrate}
	P ( z ) \propto \frac { 4 \pi d _ { C } ^ { 2 } ( z ) R ( z ) } { H ( z ) ( 1 + z ) },
	\ee
	where $d _ { C } = \int _ { 0 } ^ { z } 1 / H ( z ) d z$ is the comoving distance, $H ( z )$ is the Hubble parameter. Here we set $R ( z ) = 1 + 2 z \text { for } z \leqslant 1 , R ( z ) = ( 15 - 3 z ) / 4 \text { for } 1 < z < 5$, and $R ( z ) = 0 \text { otherwise }$. We multiply this event rate by the observed probability as the new event rate. The observed probability is gotten from a simulation of 10000 events at every point of redshift. We deem BNS events “GW detectable” when the network of ET has SNR$>8$.
	
	\item We take the sky positions from a homogeneously distribution on the celestial sphere.
	\item We  use fiducial $\Lambda$CDM to calculate the luminosity distances.
	\item The distance modulus are given by Eq.~(\ref{distancemodulus}).
	\item We use Eq.~(\ref{modelA}) to calculate the anisotropic  distance modulus $\mu_\text{diople}$.
	\item By making use of the $\sigma_{d_L}/d_L\sim d_L$ curve, we obtain the uncertainties of the luminosity distances.
	\item The uncertainties of anisotropic  distance modulus can be gotten from \be
	\sigma_{\mu} = \frac{5 \sigma_{d_L}}{\ln 10 d_L}~.
	\ee
	\item We re-sample the every distance modulus from Gaussion distribution $\mu_\text{sim}\sim \mathcal{G}(\mu_\text{diople},\sigma_{\mu})$
	\item By making use of $\{\text{(ra,dec)}, \mu_\text{sim},\mu_{\Lambda\text{CDM}},\sigma_\mu\}$, we constrain the anisotropic parameters.
\end{enumerate}

In this way, we get 100 sets of data. Each set contains five quantities: $\mu_{\Lambda\text{CDM}}, l', b', \mu$, and $\sigma_{\mu}$. Then these 100 sets of data could be used to fit the anisotropic model using the least-$\chi^2$ method. The center values of ($d,l, b$)  and uncertainties should be obtained.

We repeat the above procedures 800 times, and then calculate the average of uncertainties of ($d,l, b$).
For the cases of 200, 300, ..., and 1000 BNS merging events, the calculation method is the same as the case of 100 BNS merging events.
Our results are presented in Fig. \ref{result-13} to Fig.~\ref{result-16}.
\begin{center}
	\includegraphics[width=\columnwidth]{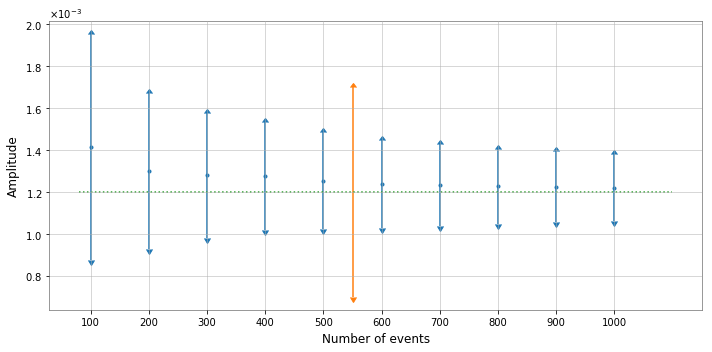}
	\figcaption{\label{result-13} The relationship between the number of BNS merging events observed by ET and the anisotropic amplitude. The uncertainty of the amplitudes are also plotted.}
\end{center}


Fig.~\ref{result-13} shows the relationship between the number of BNS merging events observed by ET and the inferred anisotropic amplitude. The uncertainties of the amplitudes are also plotted. The horizontal axis is the number of BNS merging events assumed to be observed. The vertical axis is the inferred amplitudes of anisotropic parameter $d$ and their uncertainties. It can be seen that when the number of BNS merging events observed increase, the uncertainty of the inferred amplitude becomes smaller. When 200 BNS merging events are observed in redshift $(0,1)$ by ET, the determination of the amplitude $d$ will be able to achieve the accuracy of Union2.1~\cite{Lin:2016jqp}. 
When the number of BNS merging events becomes larger, the incline of uncertainty reduction is gentle.

\begin{center}
	\includegraphics[width=\columnwidth]{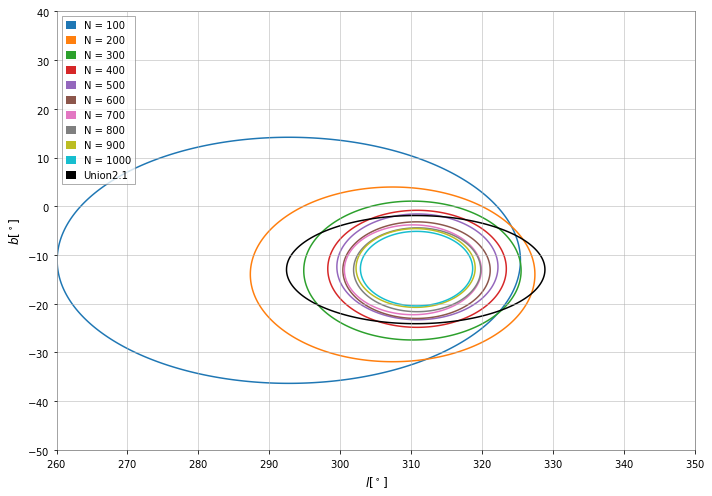}
	\figcaption{\label{result-14}  The average 1$\sigma$ confidence region in the ($l,b$) plane for different number of GW events observed by ET.
		The 1$\sigma$ confidence region of Union2.1 is also plotted for comparison.}
\end{center}

Fig.~\ref{result-14} shows the average 1$\sigma$ confidence region in the ($l,b$) plane for different number of GW events observed by ET.
The 1$\sigma$ confidence region of Union2.1 is also plotted for comparison. The different colour ellipses  represent the anisotropic direction inferred with different number of GW events. It can be seen that the inferred direction becomes more and more accurate as the number of observed BNS merging events increases. When BNS merging events observed by ET reaches 400, the determination of the anisotropic direction will be able to achieve the accuracy of Union2.1~\cite{Lin:2016jqp}.

\begin{center}
	\includegraphics[width=\columnwidth]{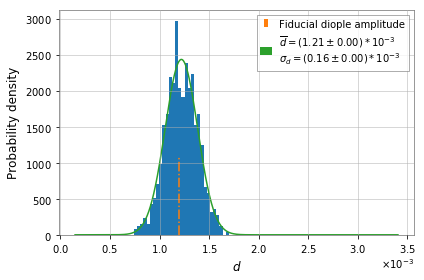}
	\figcaption{\label{result-15} Probability density map of the anisotropic amplitude inferred from 1000 simulated BNS merging events observed by ET.}
\end{center}

Fig.~\ref{result-15} shows the probability density map of the anisotropic amplitude inferred from 1000 simulated BNS merging events observed by ET.
 The distributions can be fitted by Gaussian 
 function centering at $1.21 \times 10^{-3}$, 
 with the standard deviation $\sigma_{{d}} = 0.16\times10^{-3}$. This implies that the fiducial dipole amplitude can be correctly reproduced with $\sim$ 1000 GW events

\begin{center}
	\includegraphics[width=\columnwidth]{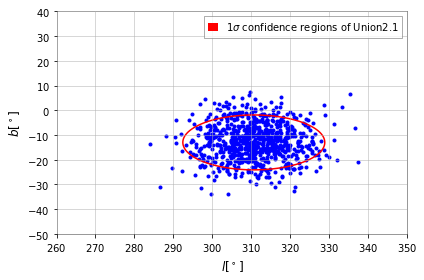}
	\figcaption{\label{result-16} The distribution of preferred directions in 800 simulations with 1000 GW events observed by ET in each
		simulation. The red error ellipse is the 1$\sigma$  confidence region of Union2.1. }
\end{center}

Fig.~\ref{result-16} shows the distribution of preferred directions in 800 simulations with 1000 GW events observed by ET in each simulation. The red error ellipse is the 1$\sigma$  confidence region of Union2.1. 
The percentage of points within the red circle is 79.2\% in Fig.~\ref{result-16} .

\section{ Discussion}
\label{sec:conclusion}
\vspace{0.1cm}

We used a new method to constrain anisotropy of space-time by using GW. Unlike the Fisher Matrix method, we simulated the process of extracting GW parameters from GW signals. By making use of GW detectors ET, we constrained the anisotropy of space-time.
200 BNS merging events in redshift $(0,1)$ observed by ET guarantee the accuracy of the constrained amplitude of the anisotropy of space-time as that from Union2.1. 

There are still some issues needed further discussions.
Weak lensing effect has not been considered in our analysis, because we don't know how weak lensing affect the waveform of GW. 
Although Planck 2018 result has been released~\cite{Aghanim:2018eyx}, it is expected that the result obtained from Planck 2015 is very close to that from Planck 2018. 

In the future, we should not infer only the four parameters but also try to include the angle of angular momentum of our line of sight, $\iota$, by using BNS merging events. Studies have shown that~\cite{Nakar:2007yr}, when $- 20^\circ<\iota<20^\circ$, one will also be able to see the electromagnetic signal.



\acknowledgments{We are greatly appreciate Yong Zhou for useful discussions.}

\vspace{10mm}

\vspace{-1mm}
\centerline{\rule{80mm}{0.1pt}}

\end{multicols}

\clearpage

\end{document}